# SmartSantander: IoT Experimentation over a Smart City Testbed

Luis Sanchez[a], Luis Muñoz[a], Jose Antonio Galache[a], Pablo Sotres[a], Juan R. Santana[a], Veronica Gutierrez[a], Rajiv Ramdhany[b], Alex Gluhak[c], Srdjan Krco[d], Evangelos Theodoridis[e], Dennis Pfisterer[f]

[a]*University of Cantabria, Plaza de la Ciencia s/n, Santander, 39005, Spain*

[b]*Lancaster University, South Drive, Lancaster, LA1 4WA, United Kingdom*

[c]*University of Surrey, Faculty of Engineering & Physical Sciences, Guildford, GU2 7XH, United Kingdom*

[d]*Ericsson Serbia, Milentija Popovica 5a/V, Belgrade, 11070, Serbia*

[e]*Research Academic Computer Technology Institute, Patras, 26500, Greece*

[f]*University of Lübeck, Ratzeburger Allee 160, Lübeck, 23562, Germany*

*Email addresses*: lsanchez@tlmat.unican.es (Luis Sanchez), luis@tlmat.unican.es (Luis Muñoz), jgalache@tlmat.unican.es (Jose Antonio Galache), psotres@tlmat.unican.es (Pablo Sotres), jrsantana@tlmat.unican.es (Juan Ramón Santana), veronica@tlmat.unican.es (Veronica Gutierrez), r.ramdhany@lancaster.ac.uk (Rajiv Ramdhany), A.Gluhak@surrey.ac.uk (Alex Gluhak), srdjan.krco@ericsson.com (Srdjan Krco), theodori@cti.gr (Evangelos Theodoridis), pfisterer@itm.uni-luebeck.de (Dennis Pfisterer)

**Abstract**

This paper describes the deployment and experimentation architecture of the Internet of Things experimentation facility being deployed at Santander city. The facility is implemented within the SmartSantander project, one of the projects of the Future Internet Research and Experimentation initiative of the European Commission and represents a unique in the world city-scale experimental research facility. Additionally, this facility supports typical applications and services of a smart city. Tangible results are expected to influence the definition and specification of Future Internet architecture design from viewpoints of Internet of Things and Internet of Services. The facility comprises a large number of Internet of Things devices deployed in several urban scenarios which will be federated into a single testbed. In this paper the deployment being carried out at the main location, namely Santander city, is described. Besides presenting the current deployment, in this article the main insights in terms of the architectural design of a large-scale IoT testbed are presented as well. Furthermore, solutions adopted for implementation of the different components addressing the required testbed functionalities are also sketched out. The IoT experimentation facility described in this paper is conceived to provide a suitable platform for large scale experimentation and evaluation of IoT concepts under real-life conditions.

**Corresponding author:**
Luis Sanchez
Universidad de Cantabria. Laboratorios de I+D de Telecomunicacion. 39005. Santander. Spain.
E-mail: lsanchez@tlmat.unican.es
Phone: +34 942200914
Fax: +34 942201488

## 1. INTRODUCTION

The Internet of Things (IoT) has recently risen in prominence due to significant advances in enabling device-technologies, such as Radio Frequency Identification (RFID) tags and readers, Near Field Communication (NFC) devices, and embedded sensor/actuator nodes. With this emergence of interconnected devices and services, the IoT has been touted to become the next major extension to the current fixed and mobile networking infrastructures. Recent predictions [1] foresee that IoT will form an essential part of the Future Internet (FI), as its connected devices will outnumber the computers and mobile devices utilised by human users by orders of magnitude. If such a scenario unfolds, it is not hard to conclude that the design of the FI and its architecture will be strongly influenced by the requirements of the IoT.

However, the IoT has many facets and exceeds the scope of currently-available deployments mainly due to two issues. Firstly, current IoT-like deployments are essentially closed and vertically-integrated solutions tailored to specific application domains. Secondly, new technologies and solution-optimisations are constrained in terms of applicability to the context under which they have been tested. For example, the research on one of the predominant areas of IoT, namely Wireless Sensor Networks (WSN), the experimentally-driven one in particular, has primarily focused on advances within WSN islands, providing optimized solutions for the resource-constrained devices of which they are composed.

Realising the vision of the IoT, therefore, requires an agreed architectural reference model, based on open protocol solutions and key enabling services that enable interoperability of deployed IoT resources across different application domains and contribute to horizontal re-use of the deployed infrastructure [2][3]. Additionally, a major goal of IoT research is to integrate WSN into a globally interconnected infrastructure, moving from the currently existing Intra-net to a real Inter-net of Things [4].

Based on this precept, the SmartSantander project [5] mainly targets the creation of a European experimental test facility for the research and experimentation of architectures, key enabling technologies, services and applications for the IoT in the context of a smart city. This facility aims to leverage key IoT-enabling technologies and to provide the research community with a unique-in-the-world platform for large scale IoT experimentation and evaluation under real-world operational conditions. Setting an experimental facility into a city context has special significance for IoT research for three main reasons: *1)* the *pervasiveness* of IoT-based technologies that form part of the Smart City infrastructure fabric and the *realism* of experimentation achieved through their use; *2)* the infrastructural *scale* and *heterogeneity* (devices, protocols and services), and the population of users that are key enablers for a broad range of experimentation; *3)* the *diversity* of problems and application domains in dense techno-social eco-systems such as Smart Cities that provide invaluable sources of challenging functional and non-functional requirements. As their infrastructure exhibit these properties Smart Cities provide excellent environments and are, indeed, catalysts for, IoT research.

Four contributions are presented in this paper. Firstly, this paper describes the architectural reference model for open real-world IoT experimentation facilities defined in the SmartSantander project. More specifically, it highlights the key challenges addressed in establishing an urban city-scale IoT experimentation facility and illustrates the platform usage though a representative set of implemented use cases. Secondly, as the deployment of large-scale distributed multi-purpose multi-stakeholder IoTinfrastructure is complexity-fraught and not risk-averse (often a compromise over platform capabilities, overall usefulness and cost), we regard the experience gained from our physical deployment process as another valuable contribution. In this respect, the paper provides detailed insight on the actual physical deployment of a large-scale heterogeneous IoT infrastructure over the city of Santander. The third contribution consists on presenting the solutions adopted for making the facility usable for the experimenters. The IoT experimentation support framework relies on the integration of existing components from SENSEI [6], WISEBED [7] and Telefonica Ubiquitous Sensor Networks (USN) Platform [8]. However, due to SmartSantander's unique requirements we have implemented additional mechanisms to address support for large-scale, horizontality, heterogeneity, mobility testing, as well as security, privacy, and trust. Finally, describing the different supported experimentation capabilities of the deployed facility is the last contribution presented in the paper.

The paper is organized as follows. In Section 2 related work and facilities for experimental IoT research are presented. Section 3 describes the SmartSantander platform's high-level architecture emphasizing the main requirements and testbed singularities that have been considered for the realization of the experimental facility. Section 4 provides insights on the deployed IoT infrastructure at the city of Santander. The mechanisms that have been implemented for the testbed management in terms of resource discovery and testbed monitoring are described in section 5. Section 6 presents the solutions developed for supporting the experimentation on top of the SmartSantander infrastructure. Finally, section 7 concludes the paper presenting some of the work to be accomplished in the near future.

## 2. RELATED WORK

Despite significant technological advances, difficulties associated with the evaluation of IoT solutions under realistic conditions in real world experimental deployments still hamper their maturation and significant roll out. The use of experimental facilities is considered a key enabler to facilitate the design and evaluation of novel IoT systems that work more reliably under realistic operational conditions and for their evaluation. A plethora of testbeds have emerged in the past decade. Many of these are lab-based testbed which suffer from various shortcomings such as realism of experimentation environment, limitations of scale and mobility testing support, heterogeneity of underlying experimentation substrate or the lack end user involvement in IoT experimentation. The reader is referred to [9] for a more detailed survey and analysis of these testbeds. Our work aims to overcome several of these shortcomings and provide a facility for experimentation with IoT deployments in urban environments and SmartCity services and applications that can be enabled on top of these.

Existing efforts that most closely match our target environment are smart city deployments such as Oulu Smart City (outdoor sensor nodes) [10] or CitySense (embedded PCs with WiFi interfaces deployed on lamp posts) [11]. Although they offer IoT devices for service enablement (Oulu Smart City) or experiments (Citysense), they do not adequately provide provisions addressing experimentation requirements such as IoT device heterogeneity, support of realistic mobility scenarios and lack adequate scale necessary for carrying out large experiments or user trials. Furthermore they are not designed with the intent to serve both as service provisioning and experimentation infrastructures.

Some of these aforementioned requirements are partially tackled in lab based testbeds. For example the KanseiGeni [12], SensLab [13] and iLab.t [14] testbeds provide adequate heterogeneity by offering different mote platforms at the IoT tier and GW tier (KanseiGeni) devices for experimentation. However, the target deployment environment differs from urban outdoor environments, so do the underlying tools or mechanisms that have been designed to manage these. Although the scale of these testbeds is significant for indoor testbeds, a city deployment can easily exceed these numbers by an order of magnitude.

Similarly WISEBED [15] offers large IoT device heterogeneity by providing support for testbed federation. In fact our framework builds upon WISEBED and its underlying capabilities and extends these for the use in a larger scale out-door environment. For example our work adds support for wireless reprogramming of experimentation nodes, improved usability for experimenters for selection adequate experimentation resources and increased robustness and lower configuration overhead for management of testbed nodes.

Unlike the other testbeds, our testbed provides access to mobile experimentation nodes that are embedded in real urban infrastructures, e.g. busses or public service vehicles, in order to allow more realistic mobility experiments. Furthermore our testbed has the ability to involve real citizens into the experimentation life cycle.

# 3. IoT Testbed Requirements and Architecture

This section, first, elaborates on the requirements for providing a rich IoT experimentation environment and addressing many open research challenges in the area of IoT testbeds. Based on these requirements, it provides an overview of the architecture of the SmartSantander testbed and the features of the platform.

## 3.1 Design considerations

As reported in previous work [9], existing network testbed facilities have several limitations that make them fail to provide adequate support for the emerging requirements of experimental IoT research. The SmartSantander facility offers a variety of properties and features to overcome many of these shortcomings and integrates them into a holistic experimentation environment. In the following we highlight the key requirements along multiple dimensions and provide considerations on how the SmartSantander facility addresses them.

*Experimentation realism:* Live testbeds provide a degree of experimentation realism that even the most detailed simulation cannot achieve [16]. We argue that, for IoT-technology experimentation, even lab-based testbeds do not suffice to evaluate research prototypes under realistic conditions and to facilitate their transfer into real world deployments. IoT technologies are heavily dependent on ambient environmental conditions in which they are deployed, including the service logic of the diverse IoT applications. Smart cities in particular are an important emerging domain for the IoT in which a multitude of application areas intersect and therefore represent a realistic/fertile experimentation medium for IoT technologies. To this end, the SmartSantander facility consists of an urban deployment within the city of Santander and other partner sites. This enables more realistic experimentation and faster maturation of IoT solutions for the mass market.

*Scale:* Real-world experimentation in a target deployment environment also requires experimentation at adequate scale. While smaller-scale testbeds with populations of tens up to hundreds of nodes were sufficient for most WSN experiments, many IoT experiments demand an order of magnitude larger scale. In order to facilitate experimentation at scale SmartSantander offers access to thousands of IoT experimentation nodes, which can be utilised for advanced experimentation scenarios.

*Heterogeneity*: Future Internets of Things will consist of a wide variety of devices integrated with other FI infrastructure and service provisioning platforms. For reasons of applicability, it is expected that the development and evaluation of protocols and other IoT technologies be undertaken under conditions that is representative of the degree of heterogeneity inherent in the Internet of Things. In this respect, the SmartSantander provides a multi-tier architecture that encompasses the most relevant device tiers of IoT systems. The IoT device tier, in particular, offers a diverse set of heterogeneous IoT nodes (sensors, actuators, QR and NFC tags and mobile-phone-based sensing-platforms) connected via different network technologies, with different mobility (fixed or mobile), and with different sensing/actuation modalities.

*Mobility:* The IoT is composed of fixed and mobile devices which can also interact with each other in real life scenarios. While some indoor testbeds offer robot-controlled mobility, it is often difficult to reproduce real life mobility patterns in such testbeds. SmartSantander therefore provides support for realistic mobility by deploying a part of the infrastructure on moving real world entities, such as buses, public service vehicles or taxis. Furthermore the mobility of users is opportunistically leveraged by allowing the smartphone of a citizen to report information captured in a participatory manner [17].

*User support and end user involvement:* Unlike many IoT and FI testbeds that are geared towards supporting the experimental researcher as its main target user, the SmartSantander facility has taken a broader approach. The deployment of such a facility in the heart of a city and the considerable costs involved motivate the exploitation of the facility beyond the experimental research community. The facility has therefore been conceived not only to act as a testbed for research with IoT technologies but for the development and evaluation of IoT enabled Smart City services and applications targeting developers of commercial Smart City services and applications. Furthermore, SmartSantander also targets end users by providing IoT enabled services to the citizens of Santander and to other beneficiaries at the different testbed sites. The involvement of concrete end users adds another dimension to the evaluation capabilities of the platform by allowing not only the assessment of technical performance of IoT solutions, but also their user adoption and social impact.

*Reliability*: Having in mind the purpose of the infrastructure, in particular that it is intended to be used for service provision, reliability of the complete system represents an important requirement to ensure smooth and uninterrupted operation.

### 3.2 ARCHITECTURE OVERVIEW

The SmartSantander platform follows a three-tiered architecture consisting of an *IoT device tier*, an *IoT gateway (GW) tier* and *server tier*. Figure 1 illustrates the three tiers representing different classes of devices and services that comprise the SmartSantander infrastructure.

The *IoT node tier* provides the necessary experimentation substrate consisting of IoT devices. These devices are typically resource-constrained (in terms of power, memory and energy availability) and export sensing or actuating capabilities. This tier accounts for the majority of the devices utilised in the testbed. Due to their outdoor deployment these devices are subject to harsh environmental conditions (physical damage, weather influences, power supply). To ensure their reliable operation, a number of measures have been undertaken. These devices are deployed at hard-to-reach locations to minimise damage from vandalism. For dependability, dual power supplies (electric distribution network combined with batteries) and dual communication interfaces are installed. For reliability, multiple communication paths to a gateway are enabled for sensor reading collection and for maintenance (e.g. over-the-air firmware and application updates) and a set of management procedures is implemented to ensure rapid detection of malfunctioning nodes.

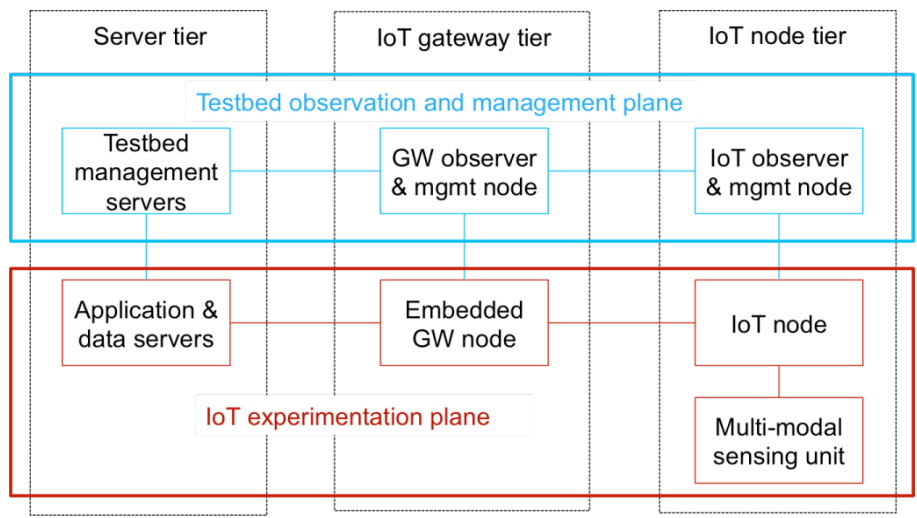

**Figure 1. Logical separation of 3-tier node architecture into a testbed observation and management and an experimentation plane**

The *IoT gateway node tier* links the IoT devices at the edges of the network to a core network infrastructure. The nodes of the GW tier are also part of the programmable experimentation substrate, in order to allow experimentation for different inter-working and integration solutions of IoT devices with the network elements of a current or FI. The GW tier devices are typically more powerful than IoT nodes but at the same can still be based on embedded device architectures – and are thus more resource-constrained than devices of the server tier.

The *server tier provides* more powerful server devices which are directly connected to the core network infrastructure. The servers can be used to host IoT data repositories and application servers that can be configured to realise a variety of different IoT services and applications or to investigate approaches for real world data mining and knowledge engineering. The server tier benefits from virtualisation in a cloud infrastructure, ensuring high reliability and availability of all components and services.

The proposed architecture is agnostic to the communication technologies between the different elements at the different tiers. In this sense, realizations of the architecture can be carried out using different communication technologies between servers, GW nodes and IoT nodes. The communication solutions adopted for the Santander testbed, for instance, are described in Section 4.2.

A key design consideration is to minimise the required human intervention to make both use and management of such large scale infrastructure tractable. Thus, the architecture has been separated into a *Testbed observation and management plane* and an *IoT experimentation plane*.

The *Testbed observation and management plane* comprises all the functionalities of the testbed dealing with dynamic management, plug-and-play configuration and automated fault management of the SmartSantander framework. A testbed user will invoke the APIs offered through the *IoT experimentation plane* in order to configure, run and control its experiments. Most experiments will utilise the nodes of the IoT node tier; however some end-to-end experiments or holistic IoT solution evaluations will require also the involvement of the gateway and server tier in the *IoT experimentation plane*.

It should be noted that this separation is logical and does not automatically imply that functionalities on different planes are hosted on different network nodes. In some cases, functions of both planes can be part of the same device while in other ones also a physical separation may exist. Physical separation has the advantage that experiments are not influenced by *testbed observation and management plane* functions, which may impair performance results on resource-constrained devices. However this comes at the cost of additional hardware.

In order to realize this architecture we propose a reference model for IoT experimentation testbeds that encompasses both testbed observation/management, and IoT experimentation planes. We contend that such facilities requires, as illustrated in Figure 2, the provision of testbed features by four main sub-systems: *1)* Authentication, Authorisation and Accounting (AAA) *2)* Testbed Management *3)* Experimental Support and *4)* Application Support. In our reference model, each subsystem comprises several functional blocks that implement the functionality expected from the subsystem. Subsystems may span across the three node-tiers requiring different components or logic to be deployed at each tier. Subsystems export a number of interfaces. Interfaces in our reference model architecture are notional entities that expose the functionality of the different subsystems through a collection of APIs. In concrete instantiations of the reference model, these interfaces may be realised through technologies such as Web Services, RESTful APIs, messaging protocols or event handling, to name but a few.

The AAA subsystem controls the access to the testbed by authenticating users, authorising the invocation of particular testbed services based on user privileges and monitoring the level of platform-use by users. Its services are exposed via the Access Control Interface (ACI).

The Testbed Management Subsystem encapsulates the functionalities concerning the automatic management of the facility. Through the exported Management Support Interface (MSI), it provides access to functions such as resource discovery, dynamic resource registration, resource or software component reconfiguration, and testbed monitoring and fault management. The MSI interface is used principally by the testbed administrator to ensure the operation of the facility.

The Experimentation Support Subsystem (ESS) embodies the experimentation plane functionality of the testbed by providing functions for testbed resource selection, specification of experiments including resource configurations, reservation of testbed resources, scheduling of experiments as well as deployment and execution control of experiments and data collection /analysis. Essentially, it provides operations to assist the user during the entire experimentation life-cycle. The ESS's functionality is exposed through the experimental support interface (ESI) which is mainly used by scientific researchers; it also possible to access the service functions of the Experimentation Support Subsystem (ESS) through this interface.

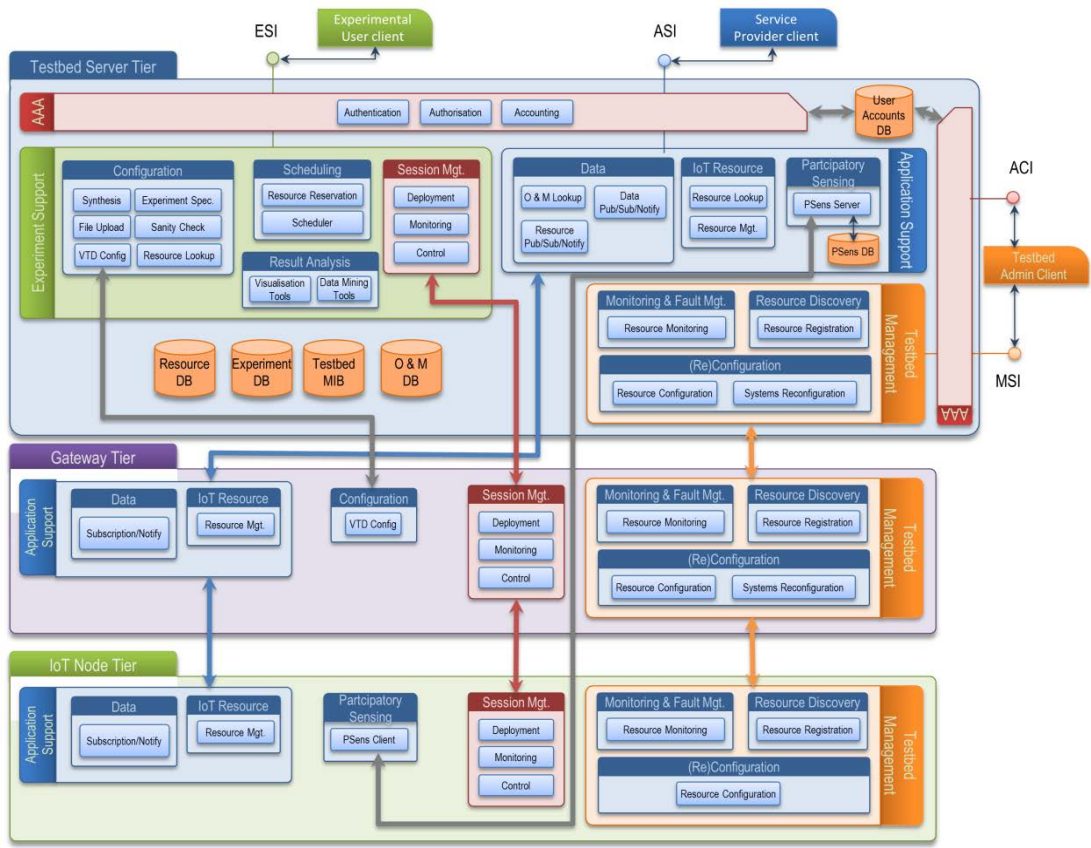

**Figure 2. Reference model architecture of the SmartSantander facility**

The Application Support Subsystem (ASS) offers via its Application Support Interface (ASI) a wide range of data management functions that can operate on information retrieved from the devices at the IoT node tier. For instance, it enables service applications to discover and select sensor data streams, issue commands to actuators, subscribe to sensor data events and access recorded sensor data for the purpose of data mining. Not only Smart City service provision will be supported through this interface but also experiments at service level that mainly needs access to the data collected within the infrastructure.

The three main functional features that have to be supported in SmartSantander, i.e. experiment support, platform management and service provision, necessitate functionality mapping and simultaneous deployment on the three architecture tiers. These three aspects have to coexist at each tier in such a way that all of them are supported but do not affect each other significantly. The only exception to this full coexistence is found at IoT node tier where some of the devices pose limits to the experimentation that can be carried out over them.

Testbed-management data and sensor-observations are stored in data repositories which can be accessed by software components from each subsystem. The repositories typically capture representations of the information models defined in our reference model. More specifically, we have defined information models for the specification of experiments, the description of testbed resources, the specification of observations and sensor readings and the specification of logical node topologies for experimentation. These information models are not presented here, as they are beyond the scope of this paper.

Our reference model reflects the experimentation/service-provisioning duality that we believe is crucial to the overall usefulness of the platform and for the definition of exploitation models to ensure its sustainability. As a concrete realisation of this reference model, the SmartSantander platform supports both experiment-execution and service applications (smart city services) concurrently within its infrastructure. Our solution for the coexistence of experiments and smart city services relies on a combination of dedicated nodes and of sensor-data sharing. Sets of dedicated nodes for experimentation and for smart city applications are required to maintain Service-Level Agreements (SLAs) brokered with the different stakeholders of the platform. Sharing of data streams from virtually every IoT node in the platform enables a multitude of experiments and smart city services to coexist as sensor-data consumers.

## 4. SANTANDER TESTBED DEPLOYMENT

The objectives of SmartSantander's deployed IoT infrastructure are two-fold as well as concurrent. As a testbed, it enables experimental assessment of cutting-edge scientific research. However, as mentioned in Section 3, this testbed goes beyond the experimental validation of novel IoT technologies. It also aims at supporting the assessment of the socio-economical acceptance of new IoT solutions and the quantification of service usability and performance with end users in the loop. For instance, it simultaneously supports the trial and subsequent provisioning of smart city services. To attract the widest interest and demonstrate the usefulness of the SmartSantander platform, the deployment of the IoT experimentation infrastructure has been undertaken to realise the most interesting and impact-generation use cases. In this respect, application areas have been selected based on their high potential impact on the citizens, thus enabling the execution of extensive experiments to obtain insights into the uptake of IoT-based services deployed in a live environment. Also taken into consideration in the selection of application use cases are the diversity, dynamics and scale of the IoT environment. All these aspects increase the potential of the testbed for the evaluation of advanced protocol solutions.

### 4.1 USE CASES AND SCENARIOS

This section outlines some of the selected use cases and scenarios that underpinned the Santander testbed deployment.

#### 4.1.1 ENVIRONMENTAL MONITORING

The current solutions for environment monitoring in urban settings usually rely on a small number of measurements stations placed at fixed locations. Although the accuracy of the measurement equipment in these units is high, their cost effectively excludes large-scale deployment to obtain measurements at finer granularity.

With the introduction of IoT technology, it is now possible to deploy a large number of low cost sensors for a fraction of cost of the current technology [18] [19]. These IoT sensors do not provide the same degree of accuracy but using a large number of measurement points and intelligent processing of the measurements it is possible to obtain sufficiently accurate measurements. In the Environmental Monitoring use case, readings gathered from fixed and mobile sensors are used as the initial indicator of the severity of the environment pollution (air quality, noise levels and luminosity levels) covering large areas. In case where conditions are observed, special alarms are generated by the system. If these observations last for long periods of time in some specific geographical region, then more accurate environment monitoring equipment is deployed. Moreover, in order to comply fully with environmental-monitoring legislation, devices offering a high-degree of accuracy are deployed temporarily at the identified pollution hotspots, thereby resulting in the coverage of a broad area at the fraction of the cost.

#### 4.1.2 OUTDOOR PARKING MANAGEMENT AND DRIVER GUIDANCE

The Outdoor Parking Management use case implies the development and deployment of a Parking Space Management service in the city of Santander. Essentially, this smart-city service enables monitoring the occupancy of outdoor parking spaces on the streets of the Santander city centre for parking-bay usage and accounting. To implement this service, ferromagnetic wireless sensors are buried under the asphalt at each bay. Peer equipment such as repeaters are deployed in an area to guarantee connectivity with the Internet such that parking occupancy information can be disseminated instantaneously to drivers, the relevant traffic control management organisations in the city or local authorities. Further, sensor data from parking bays is aggregated and used to feed parking status information to display panels located at street intersections. These data streams can be subscribed to by mobile phone applications providing, for example, navigation help to free parking spaces. Similarly, historical parking occupancy data can be analysed by municipal authorities to determine the level of parking provisioning in the city.

### 4.1.3 PARKS AND GARDENS PRECISION IRRIGATION

The Precision Irrigation use case is aimed at augmenting the automated irrigation systems currently deployed along parks and gardens to evaluate plants' requirements in water and provide for more precise on-demand irrigation. Automatic irrigation systems in use in city parks and gardens are schedule-based i.e. run preconfigured programs based on timetables irrespective of weather conditions or the water requirements of the vegetation at particular areas. Different species of shrubbery and trees have varying requirements in terms of water consumption, which is also influenced by other factors such as soil humidity. The development of WSN precision irrigation and park monitoring applications makes it easier to increase efficiency and cut down costs. IoT devices spread around the park and gardens enable agricultural data such as air temperature and humidity, soil temperature and moisture, leaf wetness and rainfall to be collected. The real-time information from the sites provides a solid base for park technicians to adjust strategies at any time. Instead of taking decisions based on some uncertain average condition, which may not be even close to reality, or having to be physically present on-site constantly, a precision park irrigation approach recognizes differences and automates management actions accordingly.

### 4.1.4 AUGMENTED REALITY

The Augmented Reality use case aims at augmenting the city scape or locations in the city with IoT endpoints to provide context-sensitive information and services at these locations. This initially involves augmenting Points Of Interest (POI) in the city, for example touristic sites, shops and public spaces with NFC tags. These tags are used to expose services or information relevant to the location/context to site-visitors. As an example of this service usage, the site-visitor's mobile-phone display can be overlaid with relevant services or tourist-targeted information, depending on their location or direction of vision. For instance, the augmented reality use case provides tourists with a "stroll in the city" experience by supplying them with location-sensitive information such as description of monuments in their preferred language. Whilst NFC tags have been deployed at the various POIs in the city, we are currently envisaging a commercial exploitation of this platform capability that involves augmenting shops with NFC tags as a means for advertising sales opportunities to customers. This will provide shops with new opportunities to build and strengthen customer relationships [20].

There are several potential windfall applications that can exploit the data collected from this platform capability. Location information and visitor frequency can be used to gauge the popularity of sites and to adjust visitor-management strategies accordingly. Tags can be coupled with more advanced services such as "feedback" from the citizens to the city council.

### 4.1.5 PARTICIPATORY SENSING

In this scenario, mobile phones are used as sensors, feeding sensed physical data such as GPS coordinates, direction (compass) and environmental data such as noise or temperature to the SmartSantander platform. Users can also subscribe to services such as "the pace of the city", where they can get alerts for specific types of events currently occurring in the city. Users also can report the occurrence of such events, which will subsequently be propagated to other users that are subscribed to the respective type of events.

## 4.2 DEPLOYED INFRASTRUCTURE

The deployment of IoT devices to compose the SmartSantander infrastructure has been motivated both by requirements for 'in-situ' experimentation and by the aforementioned smart city services. This section therefore provides details in terms of hardware specifications and deployment locations of the SmartSantander's IoT devices. The deployment of the IoT devices in a natural setting presented unprecedented challenges; this report also describes the problems encountered and the solutions formulated to resolve them.

#### 4.2.1 DEPLOYMENT LOCATIONS

The IoT experimentation facility deployed in Santander has been settled on a cyclic approach with two of the planned phases already undertaken.

The objective of the first cycle of deployment was to create a meshed WSN on fixed locations that would serve as a testing environment for the experimental validation of advanced WSN-related mechanisms. The deployment also influenced by the city of Santander smart-city service requirements and strategy, focused on three geographical areas of significance to the smart-city services. To achieve the maximum possible impact to the citizens, the deployment process intentionally accomplishes a concentration of IoT devices in the city centre (a 1 $Km^2$ area). This area has the highest IoT node density in Santander and frequent usage provides insights into the acceptance of IoT-based services running in live environments.

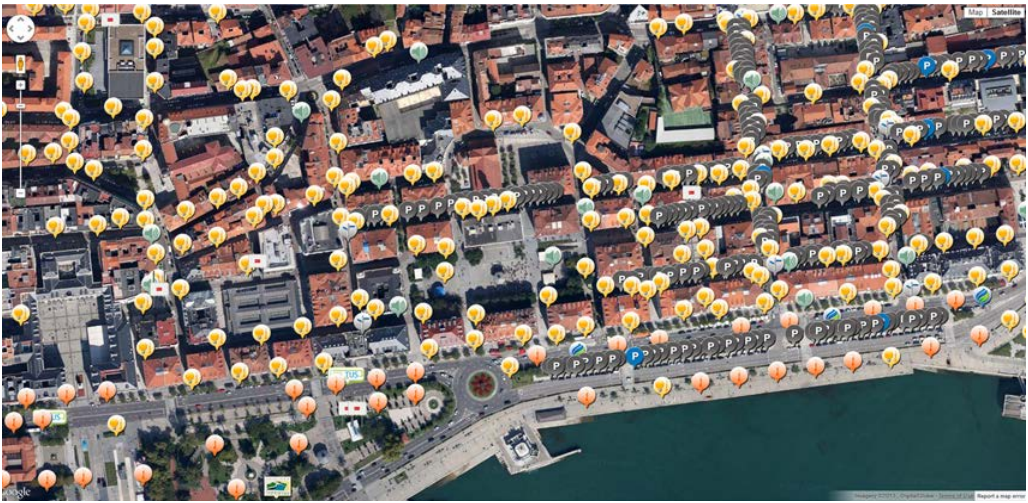

Figure 3. Santander city centre deployment excerpt view

Figure 3 shows an excerpt view of the Santander city centre deployment. The different icons represent the deployed nodes (i.e. Carbon Monoxide – CO –, light intensity, noise, temperature, and car presence detection sensors). Following the architecture described in Section 3, the deployment includes clusters of wireless sensors and gateway devices acting as cluster heads.

Once the areas for the deployment were decided, the next step in the deployment process was to specify where to physically install the devices. In this sense, the key factor influencing the decision was ensuring a viable power supply to all the devices. Although, WSNs are typically considered autonomous in terms of power needs, this assumption does not reconcile with the envisaged high-frequency multi-user usage model of our platform. Energy autonomy is achieved through the use of long-lasting batteries and most importantly, energy efficient mechanisms. However, testbed experimentation requires frequent node-software updates, which impose a stiffer power consumption penalty on IoT nodes than can be realistically met by batteries alone. To this end, WSN-experimentation testbeds such as [21], [15] or [22] rely on permanent power supplies for their nodes or exhibit a reduced node lifetime.

A hybrid solution to IoT node power requirements was adopted to minimise the infrastructure's energy consumption signature on the power grid, but ensure the survivability of its experimentation nodes. To fulfil the need for proximity to a power source, sensor devices were attached to public lampposts (as illustrated by the picture in Figure 4). The sensor devices are also endowed with rechargeable batteries and a charging circuit. Thus, daylight operation of the nodes (lampposts turned off) draws power from the batteries which are charged at night when the lampposts are turned on. Nightly operation of the nodes relies on the power from the lamppost. This solution guarantees power supply even under energy-hungry experimentation scenarios. Corresponding electrical adaptation and protections (transformer, fuse and differential protection) were added in order to obey municipal regulation.

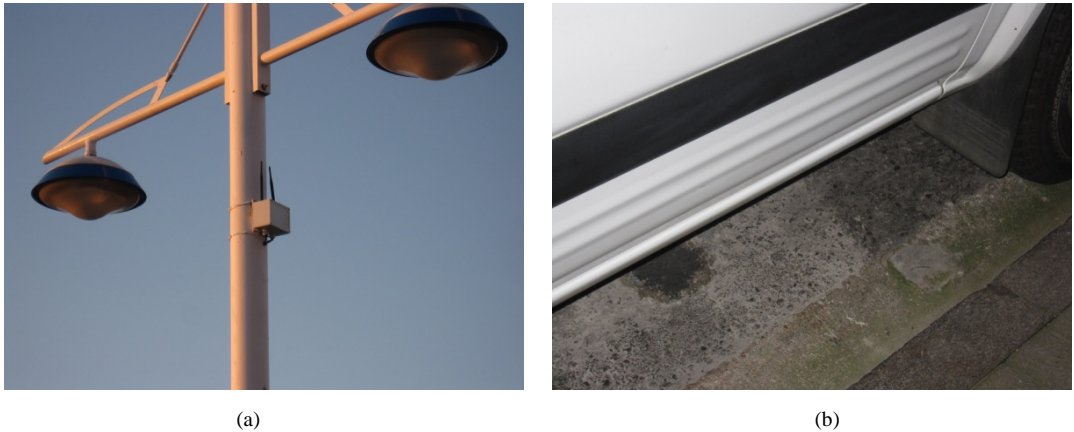

(a)  (b)

**Figure 4. (a) Wireless sensor nodes attached to lamppost; (b) Wireless sensor nodes buried under the asphalt**

Although this solution was feasible for sensor nodes supporting the environmental monitoring service, proximity to permanent power supplies for parking sensor nodes is impossible due their deployment location (buried under the asphalt, see Figure 4). Thus, due to their exclusive reliance of batteries, power consumption on these nodes is kept minimal using energy efficient mechanisms similar to those presented in commercial products like [23], [24] and [25]. This guarantees a device lifetime of over 3 years. Experimentation over these nodes is restricted only to accessing car-presence detection information.

Gateway devices have other deployment peculiarities in that they require a constant power supply and connectivity to the Internet. The solution was to install most of these devices at municipality premises located along the area to be covered. These premises are connected through a fibre-optic ring which allows GWs to be connected to a high-capacity backbone network. Where no such municipality premises were available, access to the Internet is achieved through WAN connectivity via a 3G telecoms network interface.

The first cycle of IoT deployment yielded 740 points of presence in the city. Each point of presence is equipped with several sensors making a total of more than 50 noise sensors, 600 temperature sensors, 500 light intensity sensors and 30 CO sensors. Additionally, 390 nodes with car presence detection modules have been installed in parking bays and 23 GWs have been installed to ensure connectivity between the IoT node tier and the server tier.

In the second cycle, three additional fixed-node clusters totalling approximately 50 IoT nodes were added to the infrastructure. These clusters support the smart irrigation use case and offer sensing capabilities via 45 temperature and relative humidity sensors, 25 soil moisture and soil temperature sensors, 4 weather stations with solar radiation, atmospheric pressure, anemometer and rainfall sensors, and 2 water flow sensors. The second cycle also improved node heterogeneity with the deployment of 150 mobile devices on top of public transport buses, municipality fleet vehicles and taxis. These nodes provide useful mobility patterns for experimentation as well as support environmental monitoring service. Besides the enhanced experimentation possibilities, we envisage these nodes to serve multiple application domains such as smart public transportation management and traffic conditions assessment. Further, to support experimentation based on alternative technologies and facets of the IoT paradigm, 2,000 Quick Response (QR) and NFC tags (cf. Figure 5) have been deployed over the city (at touristic POIs, bus stops and municipality's premises). These collectively support the operation of the augmented reality smart-city service.

Finally, citizens' smartphones are also part of the testbed. A Participatory Sensing mobile app has been developed within the SmartSantander project to enable these devices to send sensed physical measurements as well as mobile phone users' observations (text, images and video).

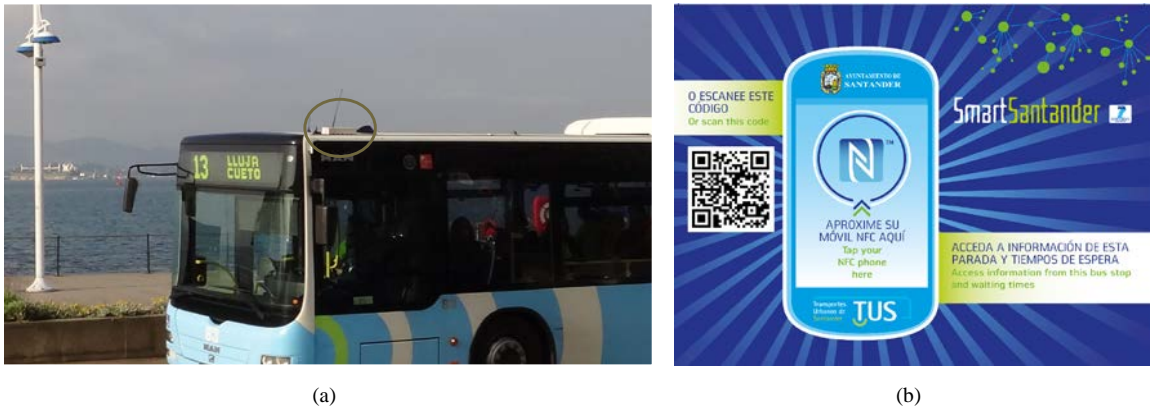

(a)                                                                 (b)

Figure 5. (a) Detail of sensor nodes installed on public bus; (b) QR/NFC tag attached to bus stop

### 4.2.2 HARDWARE DEPLOYED

Our deployment topology organises each cluster of sensor nodes around a Gateway( GW) device which provides management operations for that node-cluster and connectivity to the server tier. Where nodes are out of radio range of the Gateway device, we employ repeater nodes to ensure connectivity.

Gateway devices are intended to perform data packet routing functions so that sensor observations are transported from the sensor devices to the server tier as well as executing several experimentation and testbed management functions. Thus Gateway devices must be amply provisioned in terms of memory/processor capacity and offer communication interfaces towards both the WSN and external networks. To fulfil these requirements, embedded PCs based on the ALIX board have been used as Gateway devices. They have increased capacity in terms of processor (500MHz) and memory (256MB RAM and up to 32GB for data storage). They are configured to each include two Xbee-Pro [26] radio modules for communicating with the WSN, as well as WiFi, 3G, Bluetooth and Ethernet interfaces so that they can be connected to the rest of the SmartSantander infrastructure (i.e. SmartSantander backend and other Gateways). The GW devices run Linux OS and are encased in a housing that is IP67 compliant resistant to vandalism. The small size ensures ease of installation at appropriate ground clearance without being too conspicuous.

Sensor nodes installed on lampposts are based on the ATmega1281 microcontroller and are endowed with 8KB SRAM, 4KB EEPROM, 128KB FLASH and an extra storing SD memory with 2GB capacity. For Input/Ouput, they have 7 analogue and 8 digital interfaces available for external sensor connection, as well as 1 PWM, 2UART, 1 I2C and 1 USB interfaces. Depending on the device, the corresponding sensing probes are connected to a sensor board placed on top of the main board. This enabled the deployment of IoT nodes with diverse sensing capabilities, each with a configuration designed to support particular experiment or smart-city service classes. The sensing capabilities of our IoT devices include: air quality (temperature and CO sensors), noise (noise sensor), temperature (temperature sensor), luminosity (light and temperature sensors), irrigation monitoring sensor (temperature, relative humidity, soil moisture and soil temperature sensors) and environmental station (temperature, relative humidity, solar radiation, atmospheric pressure, anemometer and rainfall sensors).

The most noteworthy characteristic of these devices related to their wireless communication interfaces is that they are provisioned to provide two separate communication channels: one for the experimentation plane and one for the management/service plane. This is a departure from contemporary WSN testbeds [12], [15] or [21], which have traditionally relied on wired connections (e.g. USB) for supporting testbed and experiment management mechanisms. In the traditional approach, only actual experiments use the nodes' wireless interface. For instance, during a routing algorithm experiment route discovery and maintenance messages are exchanged via the wireless interface, but node reprogramming and events reporting are done through the wired interface. However, a wired backhaul for our sensor nodes is impractical in the SmartSantander platform, given the geographical distribution of sensor nodes. Making the service/management plane share the same communication channels as the experimentation plane introduces contention on the radio module and the possibility of interference between the different sources data traffic thus creating non-deterministic behaviour which is undesirable for repeatable experiments.

The solution adopted at the IoT nodes level is, as shown in Figure 6, based on the inclusion of two XBee-Pro radio modules (operating at 2.4 GHz frequency) on each lamppost sensor device. One of the modules implements native IEEE 802.15.4 protocol, whilst the other runs IEEE 802.15.4 protocol modified with the proprietary routing protocol, called Digimesh [27].

The two interfaces allow the creation of two physically independent wireless networks. The network based on the native IEEE 802.15.4 interfaces is fully devoted to experimentation. Researchers deploying their experiments on these sensor nodes will be allowed to freely use the IEEE 802.15.4 interface for communicating with other sensor nodes within the scope of their experiment. On the other hand, the network based on the Digimesh interfaces is used for carrying, to and from the cluster gateway, service provision related information as well as for supporting all the testbed and experiment management mechanisms. The reason for using a low data rate interface for this second network is mainly for guaranteeing low power consumption on the nodes.

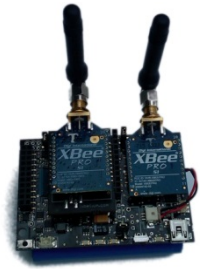

Figure 6. IoT Node deployed in SmartSantander

IoT nodes that are installed on vehicles are also equipped with a native IEEE 802.15.4 interface that can be freely used within the scope of an experiment to communicate not only among other devices deployed on vehicles, but also with devices installed on lampposts. However, since vehicles are moving all over the city, the backhaul network for service provision and management mechanisms handling is based on General Packet Radio Service (GPRS). Power consumption is not that critical for these nodes as they are powered through the vehicle batteries which represent a large energy supply for this kind of device. These devices are equipped with sensors for detecting air pollutants such as Nitrogen Dioxide ($NO_2$), CO, Ozone ($O_3$) as well as detection of particles in suspension, temperature and air humidity. Most importantly, they are also equipped with GPS so that all their observations come geo-localized and they also report speed and course of the vehicle.

Finally, for the participatory sensing use case, citizens smartphones are used as yet another IoT device. In this sense, by means of an App developed within the SmartSantander project, device sensing capabilities (i.e. GPS, acceleration, microphone, etc.) are exploited. However, what is more interesting is that through the same App, users are able to report events happening on the city (e.g. hole in the street, malfunctioning street light, full waste basket, unattended taxi stop, etc.) participating in observing the city conditions.

### 4.2.3 TESTBED INTER-TIER CONNECTIVITY

Inter-tier connectivity in the SmartSantander testbed (the Santander WSN) is arranged through different communication technologies. This section describes the network topology of the facility. As illustrated by Figure 7, fixed IoT nodes are organised into clusters that form a mesh network of nodes providing both single-hop connectivity (via the native 802.15.4 interface) and as well as multi-hop data transfer to the gateway and server tier (via the Digimesh-enabled radio interface). All the devices in a cluster form part of the same mesh network and may serve the experimentation plane or service plane or both. IoT nodes that are physically close but belong to different cluster groups are not part of the same mesh network and therefore cannot relay each other's sensor observations towards the servers. All sensor observations, testbed management and experiment management traffic have to be forwarded through the cluster head i.e. the gateway node. Multiple egress points for multi-home cluster groups have not been considered.

However, this hierarchical topology of cluster groups is not imposed on the selection of nodes for experimentation. As experiment traffic is transmitted via the native IEEE 802.15.4 interface, experimenters are given free rein to realise the topology they desire. The testbed does not impose any restriction on the use of the second radio interface. The only condition that has to be fulfilled for two of these IoT nodes to be able to communicate is the existence of a radio link between them. In essence, all the devices deployed are part of the same physical network as long as it is possible to find a set of IEEE 802.15.4 links connecting, on a multi-hop manner, any pair of the deployed nodes. This fact is presented when in Figure 7 native IEEE 802.15.4 links are set between IoT nodes in different clusters.

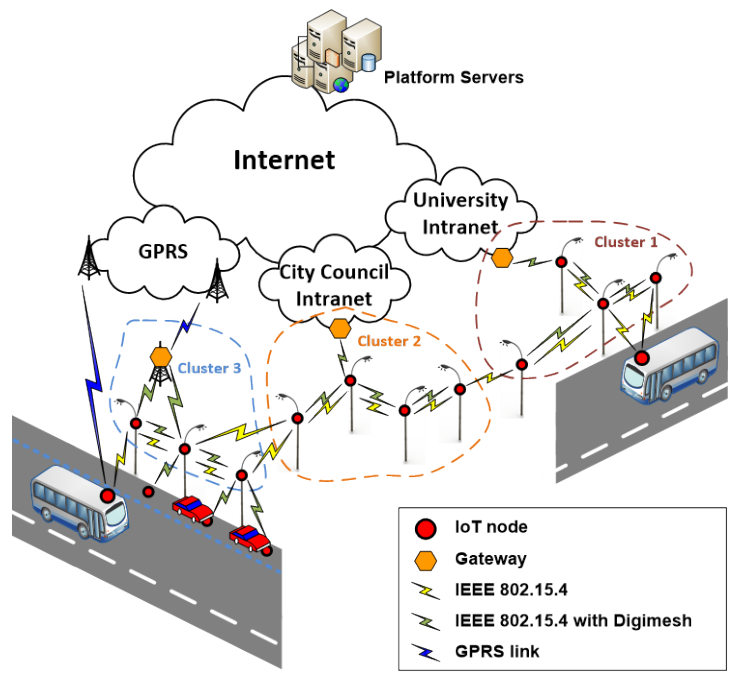

Figure 7. Testbed physical network diagram

Connectivity for the IoT nodes deployed on vehicles differs from the case of static ones. These devices are not part of any cluster but they use a GPRS connection to directly report the observations captured by their sensors and to support testbed and experimentation management procedures. However, the native IEEE 802.15.4 is capable of interacting with the fixed devices. Hence, these nodes can also be part of the abovementioned experimentation network as long as the vehicle on which they are mounted comes close to any of the fixed ones.

GWs are the cluster heads for the fixed IoT nodes. Depending on where the GW is deployed, several possibilities for connecting them to the Internet, thus to the Platform Servers, arise. Whenever it has been possible, GWs have been deployed at one of the City Council or University premises. This kind of location allows direct access to a wired Intranet. If it is not possible to find such location, GPRS connection is used to connect the GW to the core network.

Platform Servers are directly connected to the core network using the network of the University of Cantabria.

## 5. LARGE-SCALE IoT TESTBED MANAGEMENT

Beside the deployment, management of the testbed is an extremely challenging task. Developing a dependable large-scale IoT platform necessitates robust techniques for realizing out-of-band management and control planes.

Over time, there are dynamic variations to network context and to application requirements. Node membership of the network changes as new nodes are added, fail (due to power outage or hardware failure) or are disconnected (due to transient connectivity in the case of mobile nodes). Individually, each IoT node may transition through a number of possible states during the operation of the testbed; the responsiveness of a node to issued commands depends on its current state. Further, supporting multiple application domains introduces dynamic variations in the spatial and temporal characteristics of sensor data based on the new requirements of developed applications and services. Last but not least, most of the devices deployed may concurrently run experiment code from researchers while providing sensor readings for the service provision.

With the scale and variety of testbed management events to track, one cannot assume human intervention alone is sufficient to provide timely response to events and remediation to faults; a certain degree of automation is required, keeping the human in the loop only for decision-making and policy-specification. This section therefore covers features for the dynamic management of the SmartSantander testbed. Initially, the three main processes that are carried out for the testbed management are presented. Next, the components realizing them are introduced. Moreover, the resource discovery mechanisms of the SmartSantander platform, outlining the information models used for resource description and the registration process for new nodes, are presented. Finally, the monitoring feature of the testbed is described.

### 5.1 TESTBED MANAGEMENT PROCEDURES

Management processes are performed dynamically by the Management and Fault-Monitoring Subsystem, namely: resource discovery, resource monitoring and testbed reconfiguration.

The resource discovery process involves detecting new IoT resources in the testbed, registering them for use and recording the resource descriptions using standard models. Only having all the resources appropriately described using these information models to uniformly describe the attributes, capabilities and roles of the devices, experimenters or application developers will be able to select the testbed resources that best fit their needs.

The resource monitoring process concerns the dependability of the testbed platform (i.e. its robustness with respect to software component or equipment failure). IoT devices can run out of battery power, be subjected to hardware failure, accidental damage or vandalism whilst they run experiments and generate experiment traces or sensor data streams. Each experimentation node can be reserved, flashed with an experimenter's code, reset or enter an 'idle' state of service-observations reporting. Ensuring the correct execution of the IoT testbed's services in the face of such dynamicity and ensuring the testbed's resilience to failures, therefore, requires continuous monitoring of the state of its IoT resources.

On the detection of hardware failures, fault-remediation strategies require that the testbed is reconfigured to omit the faulty nodes from future experimentation or service-provisioning. Reconfiguration for testbed management is not confined to only executing fault-remediation strategies. As dynamic variation in the platform execution context occurs, reconfiguration of the platform's components is required to deliver optimal performance at all times. The reconfiguration strategy usually involves changing control parameters to optimise the operation of running components and communication protocols. Parameter-based reconfiguration is also required when application requirements change; for example the temporal granularity of sensor data can be dynamically adjusted to suit the requirements of service applications.

### 5.2 COMPONENTS FOR DYNAMIC TESTBED MANAGEMENT

A number of components have been implemented to provide mechanisms for resource discovery, resource monitoring and testbed reconfiguration. As illustrated by Figure 8, these components are deployed at different tiers of the platform.

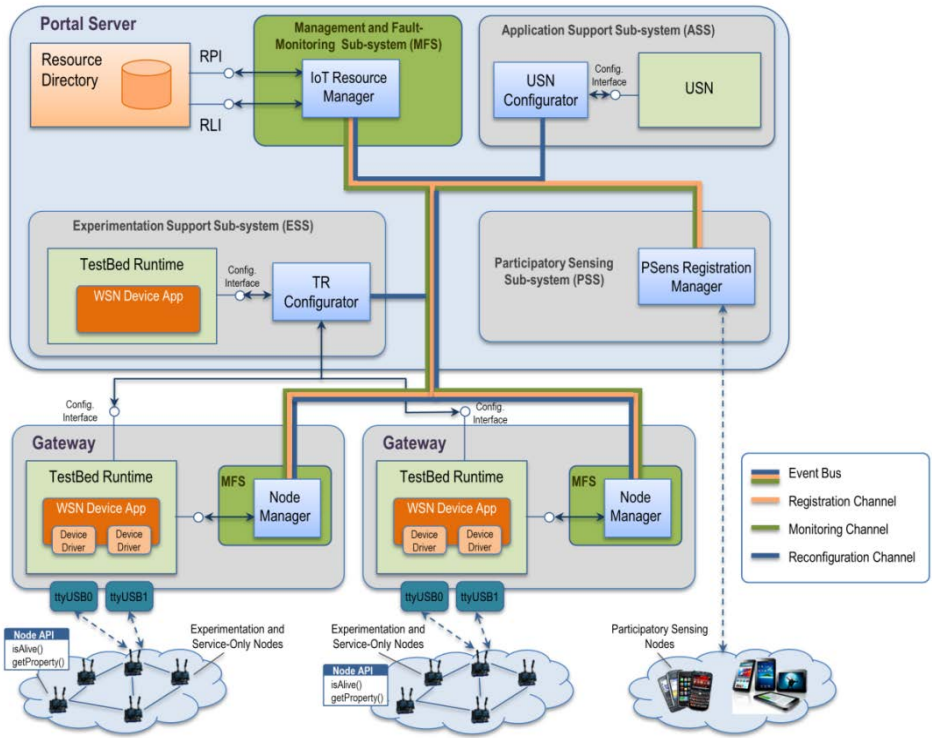

Figure 8. Components for testbed management

At the portal server level, the following components are responsible for providing functionality for testbed management.

- **Resource Directory (RD)**. It supports the resource discovery process by enabling the storage and lookup of resource descriptions for IoT nodes. It exports a Representational State Transfer (REST) interface for querying and retrieval of IoT resources based on a user's set of criteria (e.g. sensed phenomena, sensor locality, etc.).

- **IoTResourceManager**. It handles the registration of new IoT nodes in the platform and updates the status of IoT nodes based on the reception of status reports from monitoring components. It also issues reconfiguration commands to the Experiment Support Subsystem and the Application Support Subsystem based on IoT node failure detection.

- **TRConfigurator**. This component configures and controls the execution of the Testbed Runtime [7] (TR) within the Experiment Support Subsystem. It specifies the set of nodes available for experimentation to the TR in the form of a WiseML specification. It reacts to reconfiguration commands issued by the IoT Resource Manager to change the set of nodes available for reservation.

- **USNConfigurator**. This component (re)configures the USN [8] to provide services for the deployment of applications based on the set of IoT nodes (sensors and actuators) reserved for the purpose of service-provision. It adapts the resource description of IoT nodes to the SensorML format used in the USN for the purpose of node-registration. It reacts to reconfiguration commands from the IoT Resource Manager to register/unregister IoT nodes in/from the USN.

- **PSensRegistrationManager**. This component performs participatory sensing resource discovery by triggering the registration of IoT nodes such as smartphones and tablets. It is also responsible for monitoring the status of these nodes and forwarding status reports to the IoT Resource Manager.

At the gateway and IoT node tier, the following components encapsulate functionality for resource discovery, resource monitoring and reconfiguration.

- **NodeManager**. In terms of resource discovery, this component detects new nodes and triggers the registration with the RD. It also monitors the status of all nodes associated with its host GW. As periodic service message frames are routed from these nodes to the gateway, the NodeManager component intercepts the message frames to either detect new or dead nodes. NodeManager also maintains GW status using periodic beacon messages.

- **Node Application Programming Interface (API)**. A set of core function implementations from the Node API are included in every software image flashed onto the sensor nodes. They export management functions that can be invoked through command packets by the NodeManager to facilitate resource discovery and monitoring. For discovery, device-specific parameters such as the radio chip's MAC address are queried from each node by the NodeManager using the `getPropertyValue()` operation. For monitoring the integrity of each node, the `isAlive()` operation allows the NodeManager to verify the live state of each node. Status parameter-values such as a node's CPU load, memory utilisation and battery-level can also be directly queried through the `getPropertyValue()` operation.

All interactions between the management components in the Portal Server tier and GW tier occur through the propagation of events. To this end, as can be seen in Figure 8, the management plane provides dedicated event-channels for IoT resource registration, resource monitoring and testbed reconfiguration within a distributed event bus. The event bus is realized through a component, called the Event Broker (not shown in Figure 8), which embodies a generic communication substrate for disseminating management events. The Event Broker forms a distributed 'Event Bus' to which all testbed management components are connected. It implements a topic-based publish-subscribe event model wherein events are disseminated to subscribers based upon their type. The event bindings between the management components are then asynchronous, distributed and multi-party.

- *Asynchronous*: Event publishers do not block while producing events and subscribers are notified asynchronously when an event is received; this is an excellent fit for with unreliable, resource constrained WSNs.

- *Distributed*: Local or remote bindings are semantically identical, allowing components to be easily bound to local or remote event sources.

- *Multi-party*: the event bindings allow multiple consumers to be bound to the same publisher; this allows for rich interactions between components. For instance, it suffices for a management console to subscribe to the three event channels to receive information about ongoing resource registration, monitoring and reconfiguration on the platform.

The interface to the event bus is simple and lightweight. The Event Broker defines two publish-subscribe topics for each management channel, one for request events and another for the reply events. Request-reply protocols are used for each management task to instil robustness in the face of Wide Area Network (WAN) connections to remote IoT nodes.

The Event Broker uses the ActiveMQ message broker system [28] to implement the management event topics and event delivery functionality. To ensure reliable operation, features such as durable subscriptions and persistence of event topics are used from ActiveMQ. As such, events are cached for components holding durable subscriptions, should they fail or be reloaded.

Event types are implemented using Google Protocol Buffers[1] [29]; this enables the event typing system to be extensible and language-independent. The event system must first be extensible, as to enable the addition of new event types since new component-interactions can be introduced to support new platform-features. In this respect, the addition of new event types and their corresponding parser/builder functionality should be as seamless as possible. Secondly, as the platform components are developed using various technologies, bindings for the event types to different programming languages are desirable.

Table 1 summarizes the development stage and the target goal of all these components.

Table 1. Implementation status of components for testbed management

| Component name | Development stage | Target goal |
| --- | --- | --- |
| *Resource Directory* | Implemented and integrated | Thousands of resource descriptions stored. |
| *Event Broker* | Integrated | Scalable distribution of asynchronous events. Hundreds of events per minute. |
| *IoTResourceManager* | Implemented and integrated | Handling of registration and monitoring events. Hundreds of events per minute. |
| *TRConfigurator* | Implemented and integrated | Handling of registration and monitoring events. Hundreds of events per minute. |
| *USNConfigurator* | Implemented and integrated | Handling of registration and monitoring events. Hundreds of events per minute. |
| *PSensRegistrationManager* | Implemented and integrated | Dynamic registration of participatory sensing related resources. |
| *NodeManager* | Implemented and integrated | Dynamic registration and monitoring of sensor devices. |
| *Node API* | Implemented and integrated | Expose sensor devices management functionalities. |

### 5.3 RESOURCE DISCOVERY

The discovery of resources is an essential feature of an IoT platform as it serves to support selection of resources matching a user's set of criteria (e.g. sensed phenomena, sensor locality or measurement frequency).

In addition to the heterogeneity in terms of their hardware characteristics and context attributes that IoT nodes exhibit, they also differ by their intended roles within the platform. As shown by the taxonomy of SmartSantander devices illustrated by Figure 9, IoT nodes assume one of the following roles:

- **Infrastructural Nodes**: These are essentially the portal servers, gateway computers and repeater nodes that form part of the backbone network in the IoT facility, serving to run services of the testbed to support experimentation, service-provision and testbed management. Although these are mainly infrastructural nodes i.e. part, they can participate in experimentation by hosting experiment software through the use of application-sandboxing entities such as application servers, OSGI containers or virtual machines.

---

[1] The protobuf messages are included as payload into ActiveMQ BytesMessage messages. To enable event demultiplexing and handler invocation, each event-carrying BytesMessage includes an identifier as a Message Property for specifying the event type.

- **Experimentation IoT Nodes**: These are IoT nodes deployed to support experimentation. They are managed by the TR, which provides services for experimenters for node reservation and Over-The-Air Programming (OTAP) support for software deployment. They need not be exclusive to experimentation; the ubiquity of sensing for service applications often require that these nodes whilst generating experiment traces concurrently feed sensor data readings to the USN platform.

- **Service-Only IoT Nodes**: This class of nodes is reserved exclusively for the provision of services. They only deliver the observations they generate to the USN entity. These nodes cannot be reprogrammed or queried by experimenters.

- **Participatory Sensing Nodes**: These are handheld devices (for example, mobile phones and tablets) running particular Participatory Sensing applications for event tracking or collection of sensed data. They feed observations to SmartSantander's USN entity to support citizen-targeted services.

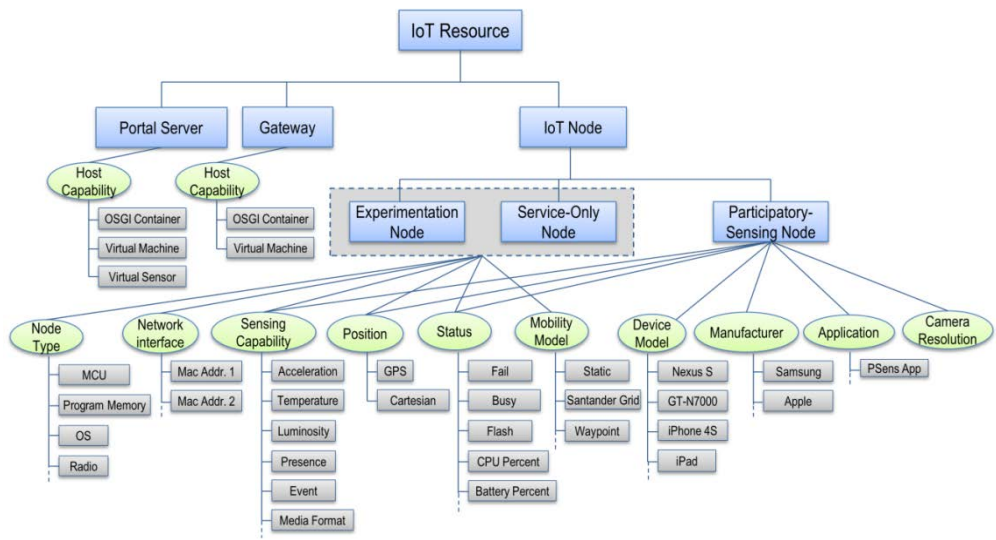

Figure 9. Taxonomy of IoT resources in the SmartSantander platform

Each class of IoT resources is described by a specific set of attributes that list hardware characteristics of the device such as node type, sensing capabilities, mobility model to name a few or time-varying context parameters such as position and device state. Resource discovery entails that these resources are searchable in terms of these attributes. The resource discovery process for the SmartSantander therefore, encompasses two main activities:

1. *Resource Management*: the specification of an IoT Resource Description Model that allows the diversity of IoT resources to be described in a consistent and uniform manner. This subsumes the storage of resource descriptions with search capabilities to facilitate lookup.

2. *Resource Registration*: the generation of resource descriptions for new IoT resources and maintenance based on the resources' dynamic state.

The two resource discovery activities are described in more details in the following sub-sections.

### 5.3.1 RESOURCE MANAGEMENT

In significantly large testbeds like SmartSantander, monitoring of nodes available in the system is one of the most important requirements for its efficient functioning. The concept of RD is often used [30] [31] in this kind of environments.

RD is an entity that stores descriptions of resources available in a system at a given time. It provides two main functions:

- Resource registration and storage of their descriptions in the RD.
- Discover of resources by searching through the stored resource descriptions.

Hence, the RD represents one of the important building blocks of the SmartSantander platform.

The initial version of SmartSantander was based on the implementation done in the FP7 SENSEI project [32]. Several extensions and modifications have been implemented on top of it to suit the new SmartSantander requirements. Further to this, the underlying Structured Query Language (SQL) database (DB) that stored resource descriptions has been replaced with a MongoDB to ensure better performance and more flexible handling of various resource description documents [33].

The RD provides two main interfaces using REST-based web services for interaction with the users and resources:

- Resource Publication Interface (RPI) allows resources to register with the RD by submitting their descriptions to the appropriate Uniform Resource Identifier (URI) of the RD. This is implemented using the POST (registration of a new resource), PUT (update of an existing resource description) and DELETE (deleting a resource description from the RD) methods of HyperText Transfer Protocol (HTTP). The resource description is submitted as a parameter of the mentioned methods.

- Resource Lookup Interface (RLI) allows resource users (applications and various platform components) to search for resources with required characteristics. This is implemented using the GET method of HTTP with appropriate set of key-value pairs as parameters of a query. RD identifies resources with the matching characteristics and responds with the list of resource descriptions. The resource descriptions contain not only a description of the resource, but also an URI that user should use to interact with the resource. Users can perform once-off lookups or can subscribe to RD asking to be informed whenever the query is satisfied. In other words, the users get informed whenever a resource with specified characteristics becomes available or ceases to be available.

Attending to the taxonomy presented in Figure 9, resources are described through Extensible Markup Language (XML) documents. Each resource description captures the main characteristics of the sensors and data they produce (type of sensors, accuracy, manufacturer, range, location, etc.) as well as the type and characteristics of the IoT nodes (testbed server, gateway, experimentation node, service only node, participatory sensing node, connection address and type, etc.).

### 5.3.2 DYNAMIC RESOURCE REGISTRATION

The dynamic registration of IoT resources involves event interactions between the NodeManager, the PSensRegistrationManager and the IoTResourceManager components. The registration of experimentation and service nodes is triggered by the NodeManager component residing at each GW. The NodeManager component produces registration request events i.e. the event set {`NODE_REG_REQUEST`, `GW_REG_REQUEST`}, and subscribes to the corresponding registration acknowledgement events i.e. the event set {`NODE_REG_REPLY`, `GW_REG_REPLY`}.

On intercepting periodic frames emitted by new nodes, the NodeManager formulates a registration request event (the `NODE_REG_REQUEST` event) and uses the interface operations of the event bus to dispatch the event on the Registration Channel (**step 1**, Figure 10). This registration event request contains the attributes required to create a valid Resource Description for the IoT node. As illustrated by the sequence of event dispatches in Figure 10, the `NODE_REG_REQUEST` event is received (**step 2**) and processed by the IoTResourceManager component to publish a Resource Description in the RD through the latter's RPI interface (**step**

**3**). This is a simplification, prior to publishing the Resource Description for the new IoT node, the IoTResourceManager performs checks to see if this node has not been seen in the platform before and verifies that the GW node is not disabled. The registrations of GWs and their associated IoT nodes are maintained as soft state through the use of timers for resource invalidation and deletion. GW nodes are responsible for sending invalidation requests for IoT sensor nodes that are no longer within its reach. They also send `HELLO` message-events to the IoTResourceManager periodically to indicate their operational status. After a number of missed `HELLO` message-events, GWs and their associated IoT nodes are first disabled (after an invalidation timeout) in the RD and subsequently deleted, should they fail to reappear after a deletion timeout.

After the Resource Description publication, the IoTResourceManager sends reconfiguration commands through the Reconfiguration Channel to the TRConfigurator and USNConfigurator components. For example, it dispatches the `ADD_SENSOR_REQ` events to the TRConfigurator component (**step 4**) to add an IoT sensor node for experimentation. The TRConfigurator receives these reconfiguration events (**step 5**), having subscribed to them and proceeds to generate the new configuration (containing the new IoT resource) for the TR. Next, it uses the configuration interface of the TR to install the new configuration (**step 6**) and upon a successful response, it sends back to the IoTResourceManager a reply event (`ADD_SENSOR_REP`, `ADD_GW_REP` or `ADD_PS_REP`) containing the result of the reconfiguration execution (**step 7** and **step 8**).

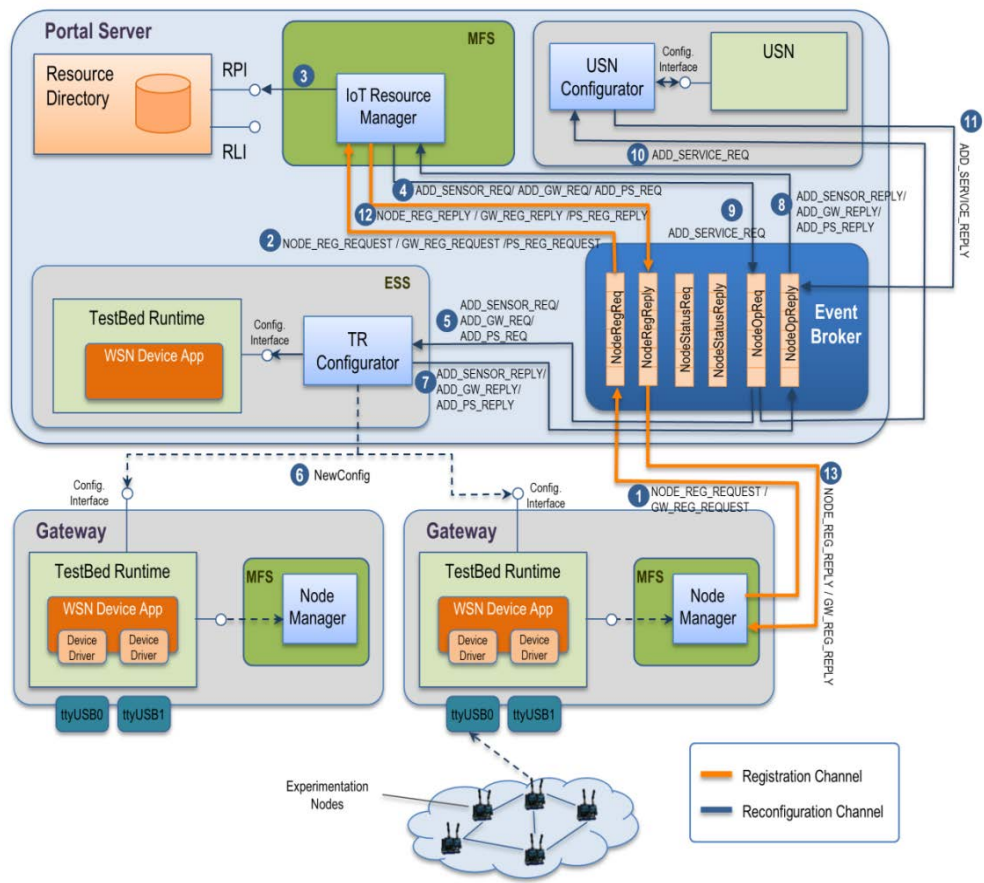

Figure 10. Event interactions for Experiment IoT Node registration

If the IoT node is destined for supporting city-services, the IoTResourceManager sends the `ADD_SERVICE_REQ` reconfiguration request event to the USNConfigurator (**step 9**). This reconfiguration request is received by the USNConfigurator (**step 10**) which effectuates a secondary registration on the USN sub-system by issuing a registration message in SensorML [34], the information model used by the USN for the description of resources. Upon the completion of this task, a reply event is sent back to the IoTResourceManager component (**step 11**) indicating the outcome of the reconfiguration request. It is only after the completion of the registration and reconfiguration tasks that the IoTResourceManager publishes a reply event to inform the NodeManager component of the outcome of its registration request (**step 12** and **step 13**). The flexibility of the event-based bindings used in the design is such that, unsuccessful registrations of IoT resources are automatically picked up by management consoles listening for the relevant events, namely, `NODE_REG_REPLY`, `GW_REG_REPLY` and `PS_REG_REPLY`.

### 5.4 TESTBED MONITORING

Due to uncontrollable factors (e.g. weather) far away from the safety of lab, testbed monitoring is very crucial for proper operation, maintenance etc. As described in section 5.1, the components for dynamic testbed management also support monitoring of resource availability and status, at three different levels of the architecture: Portal Server, GWs and IoT Nodes. As depicted in Figure 10, testbed monitoring is possible utilizing the Monitoring Channel which is established in parallel with Registration and Reconfiguration Channels. This setting permits extremely dynamic behaviour as it realize simultaneous resource registration and monitoring and appropriate testbed reconfiguration according to observations made by the other two channels.

Similarly to the Experiment IoT Node registration there are event interactions for monitoring of resources. The key component for these interactions is NodeManager. There is one instance of NodeManager running at each GW node and each NodeManager instance has the responsibility to notify the main system for the status of the corresponding GW and the attached to it IoT nodes. Periodically, NodeManager notifies with and `HELLO` event (on behalf of its GW) signalling the IoT Resource Manager that the GW is up and running. In the case that IoT Resource Manager does not receives a `HELLO` event for an already registered GW for a certain period of time (configured as parameter) then assumes that this GW is out of order and properly updates RD and reconfigures TR. Upon a fresh `HELLO` event from this GW, it is restored as active and components are reconfigured.

Furthermore, NodeManager is responsible for updating IoT Node status of nodes attached to its GW. Node Manager either by observing passing messages from the GW or by explicitly diffusing special wireless commands to IoT Nodes in the range of the GW, can extract knowledge about the status of IoT Nodes and information about them like battery level, free memory etc. Then with `NODE_STATUS_REQUEST` events Node Manager informs IoT Resource Manager about the status of the IoT Nodes. When an IoT node is not detected through passing messages or does not reacts to the special wireless commands then is considered as out of order.

ESS should be always aware for all of these changes of resource node status (GW and IoT), in order to properly reserve nodes for an experiment, execute an experiment etc. As mentioned in section 5.1, the key component for keeping up to date the ESS is TRConfigurator. TRconfigurator is the responsible for generating and maintaining the appropriate configuration state for the ESS components by transforming resource descriptions, included within the events exchanged through the Reconfiguration Channel, in the various formats (i.e. WiseML, Resource Description Framework – RDF –) that are used by the aforementioned components.

### 6. TESTBED USAGE: IOT EXPERIMENTATION

While testbed management is the most critical part for testbed administrators, the main aim of SmartSantander testbed is to be open and ready to be used by experimenters. In this sense, it is important to highlight that the experimental facility is not only heterogeneous from the point of view of the infrastructure that forms the testbed but also when looking at the kind of experimentation that is supported. In the following sections the two basic experimentation approaches that are supported are described.

### 6.1 SERVICE LEVEL EXPERIMENTATION AND SMART CITY SERVICE PROVISION

Deployment at a city scale enables direct interaction with a large base of end-users. Interaction with real end-users allows not only assessment of technologies but also assessment of services. As the environment in which the testbed is deployed is a Smart City scenario, SmartSantander aims also at experimentation at service level.

We refer to service level experimentation when experimenters make use of any of the information gathered by the deployed infrastructure in order to build a smart city application or service. The target of these applications or services is, in general, to improve the efficiency of the city and facilitating a more sustainable development of the city and its citizens.

The platform enables, through the ASI, access to any piece of information gathered by any of the deployed sensors generally following a publish/subscribe approach. The ASI also enables access to historic records of sensors matching a particular search criterion. Mostly, the queries will be related to the sensing capabilities and location of the IoT nodes.

Although the main aim of the deployed infrastructure is to enable experimenters to test their developments, the deployed infrastructure is already being used to support actual smart city services. These services are already being assessed by their corresponding end-users (e.g. municipality technicians, citizens, etc.) in several on-going trials.

Regarding the precision irrigation use case, 48 IoT nodes equipped with agricultural sensors measuring parameters like air temperature and humidity, soil temperature and moisture, atmospheric pressure, solar radiation, wind speed/direction and rainfall have been deployed.

SmartSantander developed and integrated a precision irrigation service that estimates plants' requirements in water in the different subareas of the deployment. The real-time information from the field enables park technicians to adjust irrigation strategies at any given time. Instead of taking decisions based on uncertain average conditions, which may not be even close to reality, or having to be constantly physically present on-site, a precision park irrigation approach recognizes differences and accordingly automates management actions. For this reason a smartphone application, developed for the Android platform, complements the main web application providing easy access to the measured parameters inside the park areas.

The Key Performance Indicators (KPI) evaluated during the trial mainly targeted the assessment of how realistic and accurate is the presentation of the park/garden's status and the assessment of how much the use of the IoT-supported irrigation service facilitates savings in certain resources like water and labour. Municipality technicians have helped in the assessment as they were given access to the implemented services and were asked to compare their assessment from in-field visits with the information available from the IoT-supported irrigation service. The feedback received from them was that the accuracy of the irrigation status reported through the implemented services was high enough to rely on it for taking the decision whether to water the park or not. Figure 11 (a) shows a heat-map derived from real-time measurements collected by sensors already deployed in Las Llamas Park. Similar maps and reports have been really valuable for the park managers in having a quick, remote and sufficiently accurate assessment of park status.

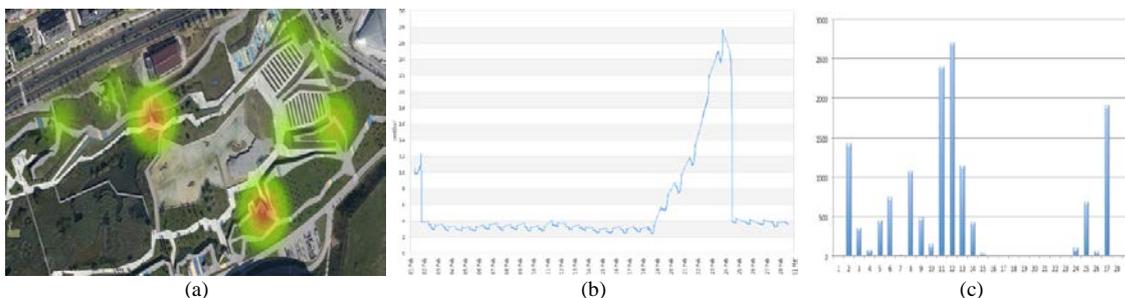

**Figure 11. (a) Soil moisture tension heatmap in Las Llamas park; (b) Soil moisture tension during February 2013 in Las Llamas park; (c) Rainfall (mm) during February 2013 in Las Llamas park**

The service also provides other details for technicians to be able to make a more in-depth evaluation. For example, Figure 11 (b) and Figure 11 (c) respectively show the soil moisture and the rainfall observed by sensors deployed in the Las Llamas Park during February 2013. The low values for the soil moisture tension, which indicates that the terrain saturated of water, shown in Figure 11 (b) fits with the rainy weather during the first two weeks of the month exposed in Figure 11 (c). However, after these first two rainy weeks there is a one-week dry period, where soil moisture increases up to 28 centibars. Nevertheless, this is not considered enough to start the irrigation system in the park. Hence, during this month, the use of IoT technology allowed parks and gardens managers to avoid visits to the park in order to do in-field inspection of the different areas and take the decision whether to irrigate or not.

## 6.2 IoT device level experimentation life cycle

In addition to service level experimentation, IoT device level experimentation is also supported by the SmartSantander testbed. Among the several differences in terms of the requirements imposed by these two kinds of experiments, one has to be highlighted. While service level experimentation does not generally need to modify the behaviour of the IoT node but just need to access the information it gathers, scientific experimentation typically needs to have complete control over the IoT device and most of the times the experiments comprises flashing the IoT node with a binary image integrating the technology/protocol/mechanism that is to be evaluated.

In this sense, a scientific experiment lifecycle has been defined for SmartSantander testbed and corresponding mechanisms have been implemented in order to address each of the different phases defined.

During specification phase, mainly dealing with the resource selection, the user is assisted with an exploration of available testbed resources and their static and dynamic properties and topological interdependencies. The user is able to formulate queries for specific resource properties in order to satisfy the requirements for a particular experimentation scenario which are matched against the testbed resources descriptions in order to provide the user with a selection of testbed resources fulfilling the desired properties. Furthermore, during setup phase, dealing with actual reservation and scheduling, ESS assures that experiments do not collide in time. Finally, on the Execution phase experimenter is empowered with experiment execution control, experiment monitoring, data collection and logging.

### 6.2.1 Resource Requirements

The SmartSantander platform aims at supporting execution of various experiments on a large scale. Each experiment involves a number of various IoT nodes, depending on the type of experiment. It is possible and preferable to have multiple experiments running at the same time, using different nodes at various or even same locations. With thousands of available nodes and multiple simultaneous requests for execution of experiments with differing requirements and involving a large number of nodes, it is necessary to provide tools and procedures for automatic assignment and scheduling of available IoT nodes to each experiment taking into consideration the capabilities of each node and requirements of experiments.

RD is used for this purpose. It contains semantic descriptions of all available resources, including information about the capabilities of the IoT nodes that are important for experimenters. It is envisaged that in a similar manner, semantic descriptions of experiments will be stored in a RD (the same or a separate instance). These descriptions will contain information about the nature of the experiment, the information needed, capabilities of the IoT nodes and supported protocols, duration; how an experiment is influencing the environment (for example one experiment might influence the outcome of another due to the activities undertaken – obstructing radio transmissions, making some information unavailable or changed) etc.

The platform will be then in position to reason over the provided semantic descriptions, matching not just an experiment with appropriate resources, but also making sure that the new experiment will not interfere with other experiments scheduled at the same time. This mechanism will greatly improve the efficiency of the allocation of resources to the experiments and will ensure proper condition for all simultaneous experiments as well as the services running on the platform.

### 6.2.2 RESOURCE RESERVATION AND PROVISIONING

ESS implementation architecture [35] was designed with generality in mind and at the architecture's core a set of standardized web service APIs allows a technology-agnostic standardized way for users to access a testbed's resources. The so-called (TR) is the reference implementation of the APIs for testbed management and experiment execution defined in the WISEBED project. It creates an overlay network for easy node addressing and message exchange independent from the actual underlying network connections.

One of these APIs is the Reservation System (RS) API, which allows users to reserve a set of resources (i.e., IoT devices) for experimentation. This API allows experimenters to select a subset of resources, uniquely identified by a Uniform Resource Name (URN), based for example on device type, attached sensors, mobility support, etc. As it has been introduced in the previous section, resource management solutions that have been put forward in SmartSantander testbed aims at improving the efficiency of the allocation of resources to the experiments and will ensure proper condition for all simultaneous experiments as well as the services running on the platform.

As can be seen in Figure 12, after successful authentication, experimenters can reserve, by sending the set of URNs to the RS web service, the devices that best fit their requirements for a certain period of time. The RS then checks authorization and reserves the devices if they are available for the desired time period. As return value the user receives a secret reservation key which he uses to access his experiment through the WSN API. This secret key is called the reservation key in the following. As a result of this invocation, users obtain a so-called secret reservation key which is used later on to identify the user as the owner of this reservation.

### 6.2.3 EXPERIMENT CONTROL

With a valid reservation key a user is able to interact with the testbed via WSN web service API. Interaction means either to control an experiment (i.e., to reprogram or to reset devices), or to interact with a running experiment (e.g., to send command or to receive benchmarking results). For both types of interactions users have full control over the complete experimental setup.

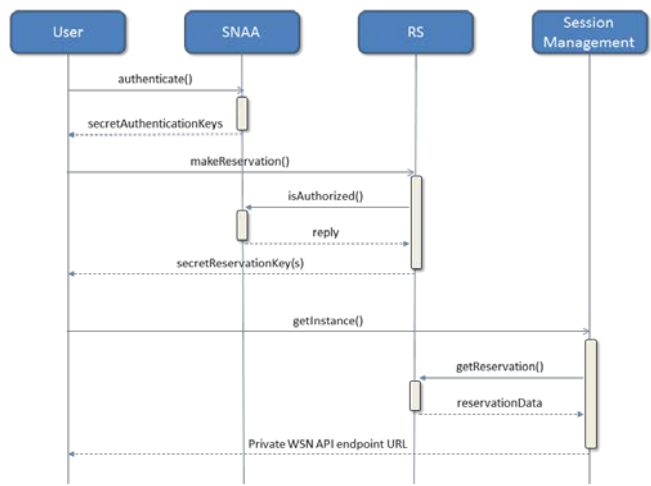

Figure 12. Interaction of the different WISEBED APIs to obtain a private testbed instance for experimentation

In order to interact with an experiment, users send their secret reservation key to the testbed via a web service call and if the experiment has started, they get a private web service endpoint URL to interact with the reserved resources. This Uniform Resource Locator (URL) points to an instance of the WSN API. The details of the full API interaction are depicted in Figure 12. For the experimenter, these steps are automated and a number of clients to these APIs are available ranging from a command-line client to web-based interface (cf. Figure 13), which both support scriptable experiments (e.g., for automatic execution of repeated experiments, etc.). To receive output generated by devices as well as status updates about the experiment, the user provides the testbed with a URL where he exposes the so-called controller API. The methods of the experimenter controller API are called by the WSN API implementation to send experiment output or asynchronous status updates on ongoing operations, like (re-)programming nodes, to the owner of the experiment.

This enables the experimenter to not only be able to control the experiment behaviour but also to transparently get the traces of the experiment so that they can be analysed on a real-time basis or stored for offline assessment and post-processing.

The overlay network created by the TR performs message forwarding and offers communication primitives that are used for the control and management of experiments. This overlay network handles the messages exchanged between the experimenter controller and the IoT Nodes enabling a virtual point-to-point connection between the experimenter host and each of the reserved nodes. Objective is to have one virtual connection per IoT Node in the testbed, accessed through an exclusive connector, enabling not only data collection or experiment logging but also experiment control operations to be triggered on a per IoT Node basis. This provides a great degree of flexibility within an experiment.

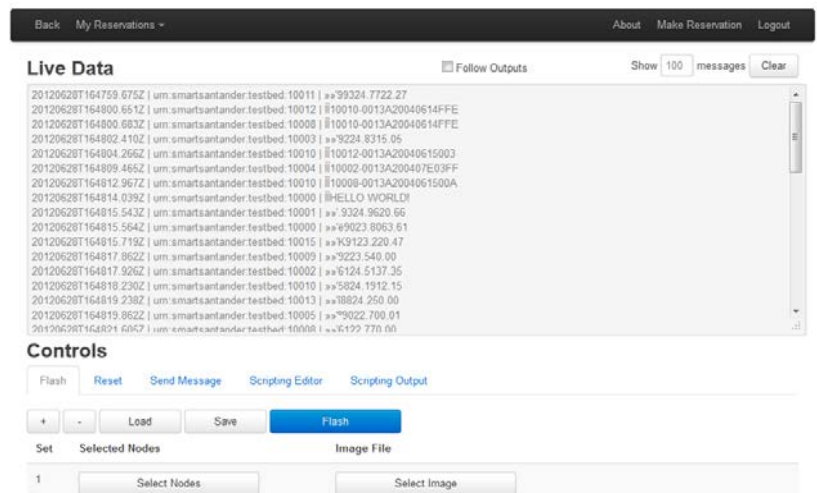

Figure 13. Web-based interface for experimentation

While achieving this might be straightforward for those testbeds relying on a wired connection to each of the IoT Nodes deployed, in the SmartSantander fully wireless context described in section 4, exclusive virtual connection with each of the testbed IoT Nodes required appropriate multiplexing and demultiplexing of the communication (both uplink and downlink) since all the virtual connections with the IoT Nodes are supported over one single physical connection (i.e. the wireless link between the GW and the IoT Nodes of his cluster). SmartSantander testbed implements at GW level the modules that generate the handlers for each of the IoT Nodes that are managed by that gateway. The TR will be able to interact with the IoT Nodes through these handlers as if they were point-to-point physical communication ports towards the underlying IoT Nodes.

### 6.2.4  IOT NODES REMOTE REPROGRAMMING

Among the different experiment control functionalities, there is one that required particular attention as it is the basis for a testbed to be considered an experimentation facility. The ability to re-program the IoT Nodes at any time during the experiment is critical and has been carefully addressed within the SmartSantander testbed.

It comes without saying that direct reprogramming of the IoT Nodes deployed in SmartSantander testbed is not a possibility. Further to this, it has been already stated that no wired infrastructure is available to support fast and resilient flashing of nodes. Thus, remote reprogramming of IoT Nodes is being carried out through OTAP mechanisms implemented as part of the ESS. Bearing in mind that clusters of IoT Nodes deployed embraces multihop wireless networks configurations, it is more precise to speak about Multihop Over The Air Programming (MOTAP).

Several MOTAP mechanisms are available in the literature [36]. However, none of them had an available implementation for the devices actually deployed. Thus, in order to make the deployed platform as dynamic and reconfigurable as possible, a reliable MOTAP protocol has been implemented for flashing nodes over the air either in unicast, multicast or broadcast fashions, as many times as needed. As it can be seen in Figure 13, flashing and resetting operations are available to the experimenter.

## 7. CONCLUSIONS

As has been indicated, Internet of Things is foreseen to be an essential part of the FI. In this paper the key features and properties that are supported by SmartSantander testbed have been described. The experimental research facility presented in this paper aims at supporting the testing of proposed protocols, services and configurations in a realistic setting at an appropriate scale. Shortcomings of the existing testbeds in terms of scale, heterogeneity, mobility and more importantly realism of experimentation environment and end-user involvement are overcome by the holistic experimentation environment deployed in SmartSantander.

This paper presents the testbed architecture as well as the main deployment issues and experiences. In this sense, specific mechanisms have been integrated in order to guarantee that the testbed is ready to provide to the experimenters all the potential that such a large scale testbed has. Particular attention has been put on the testbed management as it is of utmost importance to keep track of all the testbed resources thus guaranteeing the dependability of the facility. Moreover, the solutions developed are on their own a significant contribution that addresses the challenging tasks that are raised by the scale and variety of testbed management events to track.

The facility will be further improved and enhanced by federating it with additional sites providing access to an even larger number and more varied types of IoT devices. Additionally, the facility is being prepared for federation with other FI experimentation testbeds aligning this federation with the already existing activities in GENI [37] and FIRE [38].

### ACKNOWLEDGEMENTS

This work is funded by research project SmartSantander, under FP7-ICT-2009-5 of the 7th Framework Programme of the European Community. Authors would like to acknowledge the collaboration with the rest of partners within the consortium leading to the results presented in this paper.